\documentclass[twocolumn,showpacs,preprintnumbers,amsmath,amssymb]{revtex4}
\usepackage{graphicx}
\usepackage{dcolumn}
\usepackage{bm}
\usepackage{epsfig}
\usepackage{subfigure}

\begin{document}
\title{Dynamical Mean Field Study of Model Double-Exchange Superlattices}
\author{Chungwei Lin, Satoshi Okamoto, and Andrew J. Millis}

\affiliation{ Department of Physics, Columbia University
\\ 538West 120th Street, New York, New York. 10027}

\begin{abstract}
A theoretical study of [001] ``double exchange" superlattices
is presented. The superlattice is defined in terms of an $AB$O$_3$
perovskite crystal. Itinerant electrons  hop among the B sites
according to a nearest-neighbor tight binding model and are
coupled to classical ``core spins".  The $A$ sites contain ionic
charges arranged to form an [001] superlattice which forces a
spatial variation of the mobile electron charge on the $B$ sites.
The double-exchange interaction is treated by the dynamical mean field
approximation, while the long range Coulomb interaction is taken
into account by the Hartree approximation.  
We find the crucial parameter is the Coulomb screening length.
Different types of phases are distinguished and the interfaces
between them classified.
\end{abstract}
\maketitle


``Strongly correlated" transition metal oxides are of great current
interest because of the wide variety of novel ordered phases they
exhibit \cite{Imada 98}. A particularly striking feature is the
strong coupling between order and the ability of electrons to move
through the crystal. For example, the Goodenough-Karanori 
\cite{Goodenough, Kanamori 59} rules
establish a connection between orbital ordering and the overlap of
electron wave functions between different sites, while in
double-exchange systems such as the colossal magnetoresistance
manganites, relative spin orientation strongly couples to hopping
amplitudes \cite{Zener 50}. Very recently, experimentalists have
succeeded in fabricating high quality atomic-scale ``digital
heterostructures" consisting of combinations of correlated
materials, typically characterized by different free carrier
density and by different forms of long range order. An example of
a digital heterostructure is  [001] (LaMnO$_3)_m$(SrMnO$_3)_n$
\cite{Kawai02,Eckstein,Hwang02,Tokura02}: $m$ planes of
LaMnO$_3$ followed by $n$ planes of SrMnO$_3$, with the whole
making a periodic structure with a repeat distance of $m+n$ times
the mean Mn-Mn $c$-axis distance. Here LaMnO$_3$ (one electron per
Mn $e_g$ state) and SrMnO$_3$ (no electrons per Mn $e_g$
state) are the two end-member compounds of the `colossal'
magnetoresistance (CMR) alloy La$_{1-x}$Sr$_x$MnO$_3$  family of
compounds.

This experimental success raises fundamental questions. 
Correlated electron materials are interesting because of phases
they exhibit (for example magnetic, superconducting, and Mott insulating). 
In correlated electron heterostructures the key questions are: what
phases can occur, and what is the spatial structure; in particular
what is the character of the domain walls which separate regions
of different spatial orders? In this paper we present a detailed
study of a simple model which yields insight into
these issues.

\begin{figure}[tbp]
\vspace{0.8cm}
\includegraphics[width=0.8\columnwidth]{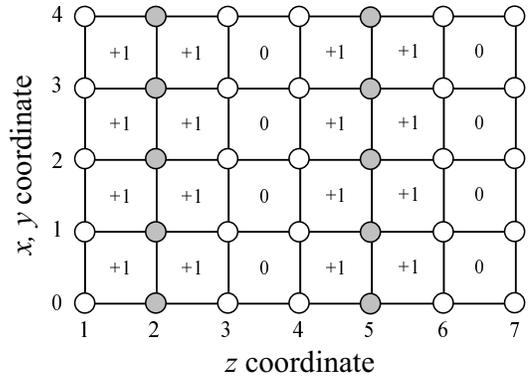}
   \caption{Schematic representation of a ($AB$O$_3)_2$($A'B$O$_3)_1$
   (2,1) superlattice counterions. $A$ and $A'$, located at 
   positions $z = (p+1/2)a$
   (with $p$ an integer), are represented by their charges +1 and 0
   respectively, whereas the two symmetry-different B sites, located at 
   integer positions, are represented by filled and open circles. }
   \label{fig:superlattice}
\end{figure}

Our model is motivated by the colossal magnetoresistance
heterostructures (CMR) now being fabricated \cite{Kawai02,Eckstein,
Hwang02,Tokura02}. It involves a
heterostructure defined electrostatically by a periodic array of
charged counter ions \cite{Satoshi, Okamoto04-2}, 
with carriers subject to the double exchange (DE)
interaction \cite{Furukawa95,Millis96} which is crucial to the physics of
the CMR materials. The lattice structure considered here is based
on the $AB$O$_3$ perovskite structure with lattice constant
$a$, and we shall be interested in structures of the form 
($AB$O$_3)_m$($A'B$O$_3)_n$ [(m,n) heterostructure]
periodic in the [001] direction. 
A schematic representation is shown in Fig.\ref{fig:superlattice} for the
(2,1) heterostructure. $A$ and $A'$ have ion charge
+1 and 0 (neutral) relative to $B$ site, and therefore total electron
density per unit cell is $m/(m+n)$. 
We place the electrically active $B$ sites in planes $z=pa$ with $p$
an integer, $m$ charge +1 ions planes planes $z = (p+1/2)a$ with $p=1$ to $m$, and $n$ charge 0
ions planes $z=(p'+1/2+m)a$ with $p'=1 ... n$. 
The conduction electron hopping between $B$ sites is described by
a one orbital tight binding model.

The Hamiltonian is
\begin{eqnarray}
H_{tot}=H_{hop}+H_{Hund}+H_{Coul}
 \label{eqn: Starting H}
\end{eqnarray}
with
\begin{eqnarray}
H_{hop}&=&-t\sum_{\langle i,j\rangle,\sigma} \left( c_{i,\sigma}^{\dag}
 c_{j,\sigma} +h.c. \right) \\
H_{Hund} &=& -J\sum_i \vec{S}\cdot
\vec{\sigma}_{\alpha, \beta}
 c_{i,\alpha}^{\dag} c_{i,\beta}
\end{eqnarray}
and
\begin{eqnarray}
H_{Coul}= \sum_{i\neq j} && \left[\frac{1}{2\varepsilon} \frac{e^2 n_i
n_j}{|\vec{r}_i-\vec{r}_j|} 
+  \frac{1}{2\varepsilon}
\frac{e^2}{|\vec{R}^{A}_i-\vec{R}^{A}_j|} 
\right. \nonumber \\  &&
 \left.  -\frac{e^2
n_i}{\varepsilon|\vec{r}_i-\vec{R}^{A}_j|} \right]
\end{eqnarray}
with $n_i=\sum_{\sigma}c_{i,\sigma}^{\dag} c_{i,\sigma}$ the occupation
number at $B$ site $\vec{r}_i$. $\vec{r}_i$ and $\vec{R}^{A}_i$
label the positions of the $B$ and $A$ sites respectively, 
and $\varepsilon$ is the
dielectric constant of the material. To solve this model, we use
dynamical mean field theory \cite{Georges,Okamoto04-2,Potthoff99}
for the double-exchange interaction and Hartree approximation for
the long range Coulomb interaction. The leading instability of the
paramagnetic phase, and also the $T=0$ phase boundaries 
are obtained by the method developed in Ref\cite{Georges,DMFT Instability}
while the non-zero $T$ phase boundaries are estimated by computing the
energy and entropy difference between different phases.
Details of calculations will be presented elsewhere.

In this model, the heterostructure is defined by Coulomb forces,
the important order is magnetic, and the coupling between order and
itineracy is via the double-exchange mechanism. However, we expect
our qualitative conclusions to apply more generally to any
situation in which the charge density varies across the
heterostructure and the physics involving a coupling between order and charge
mobility.

The model we study involves two fundamental parameters: $\alpha =
e^2/\varepsilon t a$, measuring the strength of the Coulomb
interaction relative to the electron hopping, and the Hunds
coupling $J/t$, expressing the degree to which magnetic order
controls electron occupation and hence hopping. Our results are
not very sensitive to the magnitude of $J/t$, provided it is large
enough that the conduction band is fully polarized in
ferromagnetic(FM) ground state, so we take $J/t=6$, a value
believed to be roughly consistent with the values found in the CMR
materials.

The important parameter is $\alpha$. It is sometimes convenient to
express $\alpha$ in terms of a screening length $L_{TF} \approx a/\alpha$.
At small $\alpha$, the charge
is only weakly confined. For short period structures, the charge is
uniformly distributed and the system exhibits essentially the same
phase as is found in the randomly doped bulk material. For long
period, the heterostructure has gradual charge modulation from
$n\approx 1$ ($AB$O$_3$) to $n \approx 0$ ($A'B$O$_3$). 
In this latter case, the known bulk phase diagram \cite{DMFT Instability, Dagotto 98}
leads is to expect a spatial variation of the magnetic phases,
from antiferromagnetic (AF) in the $n\approx 1$ region, to phase
separation (PS) in the intermediate transition region, and to ferromagnetic 
in the lower density $AB$O$_3$ region.
For large $\alpha$, the charge profile is more abrupt,
and the possibility of a sharp AF/FM domain wall exists. To study
this case in more detail we consider a (2,1) heterostructure which
is simple enough to study in detail and will be seen to capture a
wide range of phenomena.


\begin{figure}[tbp]
   \centering
   \includegraphics[width=0.8\columnwidth]{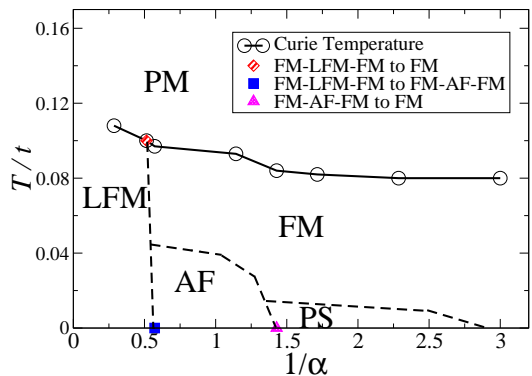}
   \caption{Phase diagram at $T-1/\alpha$ plane for $J/t=6$, (2,1) superlattice.
   PM: paramagnet, LFM: layer ferromagnet, FM: ferromagnet, AF: antiferromagnet,
   PS: phase separation.}
   \label{fig:T-alpha phase diagram}
\end{figure}

The (2,1) heterostructure has two electronic regions: a bilayer of
$B$ sites, denoted by open circles in Fig.\ref{fig:superlattice},
each with one $A$ (with charge +1) and $A'$ (with
charge 0) site as neighbor, and therefore a relatively lower
charge density; and a single layer of $B$ sites with two $A$ sites
as neighbors and therefore a relatively higher charge density. The
behavior of bilayers is found to be simple, being paramagnetic or
ferromagnetic according to the temperature. The behavior of
the mono-layer sandwiched by two $A$ layers is more complicated, 
involving also an interplay between charge binding and the nature
of the magnetic order.
Figure \ref{fig:T-alpha phase diagram} shows the calculated phase
diagram in the temperature-charge binding interaction plane, with
different phases distinguished by dashed lines. The
solid line marked by open circles indicates the Curie temperature,
below which the outer layers order ferromagnetically. Near $T_c$
the inner layer is ferromagnetic and is either aligned
($1/\alpha>0.6$) or anti-aligned ($1/\alpha <0.6$) to the outer
layers (canted phases are not found). In either case
the inner layer polarization is much smaller than that on the
outer layer. When the Coulomb interaction is weak ($1/\alpha>3$),
the charge density is weakly modulated relative to the mean value 2/3 and
ferromagnetism is observed at all $T<T_c$, consistent with the bulk 
phase diagram \cite{DMFT Instability, Dagotto 98}.

As the charge binding is increased, the central layer charge
density increases, eventually reaching values for which ferromagnetism is not
favored in the bulk phase diagram. In this $1.4< 1/\alpha < 2.8$
region, at low $T$ the central layer exhibits phase separation
between ferromagnetic and antiferromagnetic states. Phase separation
is also found in the bulk case in approximately this region,
but the phase boundaries are
slightly shifted because of a proximity effect arising from the FM
outer layers. The FM-PS phase boundaries are 
found to be second order in this
model.

As the charge binding is further increased, a 
temperature-driven first order FM/AF
transition occurs. In this $0.6 < 1/\alpha< 1.4$ region, the
central layer charge densities correspond to values at which the corresponding
bulk materials are phase separated between FM and AF states. We interpret this
FM-AF-FM phase in the superlattice as a phase separation  
in the $z$ direction: the relatively
stronger charge binding means that it is energetically favorable
for the system to phase separate by moving charge only in the $z$
direction. Finally, as the charge binding is yet further
increased ($1/\alpha>0.6$), we find a new layer ferromagnet (LFM) phase
where both central and outer layers are in-plane ferromagnetic
but with magnetizations anti-aligned. This phase is
not found in bulk calculation of single-band DE model, occurs.

\begin{figure}[tbp]
   \centering
   \includegraphics[width=0.8\columnwidth]{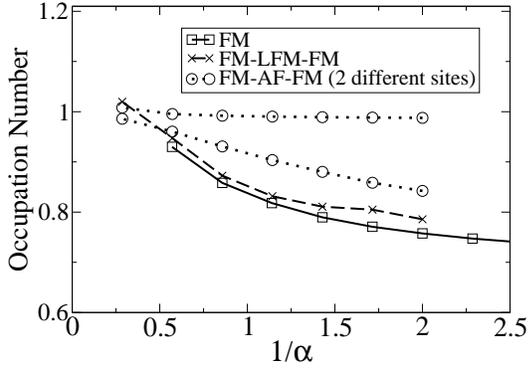}
   \caption{$\alpha$ dependence of the central-layer charge density
      for different magnetic orders. For the FM-AF-FM state, two values for
      two inequivalent sites are shown. }
   \label{fig:n(alpha)}
\end{figure}

The transitions to the AF and LFM phases are first order, and are
driven by the interplay of ordering and charge mobility.
Figure \ref{fig:n(alpha)} shows the central layer charge density as a
function of charge binding parameter, for the different
homogeneous phases (the total charge density is of course fixed
by charge neutrality). The FM phase is most favorable for electronic
itineracy, and therefore has the lowest central layer charge density. 
It thus has the least favorable Coulomb energy. The AF
phase has noticeably higher mean charge density, which moreover
exhibits the expected sublattice structure, being highest on the
sublattice with spin antiparallel to the ferromagnetic region. At
intermediate $\alpha$ the LFM phase has a lower charge density
than the AF phase, essentially because the FM core spin arrangement
results in a wider in-plane bandwidth than AF and therefore
forces more states (than AF) to be above the chemical potential.
However, at sufficiently strong charge binding the central layer
occupation becomes larger than that of the AF state, so the LFM
phase becomes favored by the Coulomb energetics.

We next discuss another general implication of our results. 
Figure \ref{fig:n(alpha)} shows that the electronic density distribution is 
strongly affected by the magnetic order changing both with dielectric
constant and with temperature. In the present model, this behavior
is a consequence of the ``double exchange" physics of coupling of
hopping amplitude to intersite spin correlations, but similar physics
may also be expected to occur in orbitally degenerate systems, where hopping 
amplitudes depend on orbital overlaps which are changed by orbital order.
This raises the intriguing possibilities of magneto-electric coupling;
for example, changing a charge density by applying a magnetic field
large enough to eliminate the antiferromagnetism or changing magnetic
behavior by "gating" the electron density\cite{Hwang 05}.

\begin{figure}[tbp]
   \centering
   \includegraphics[width=0.78\columnwidth]{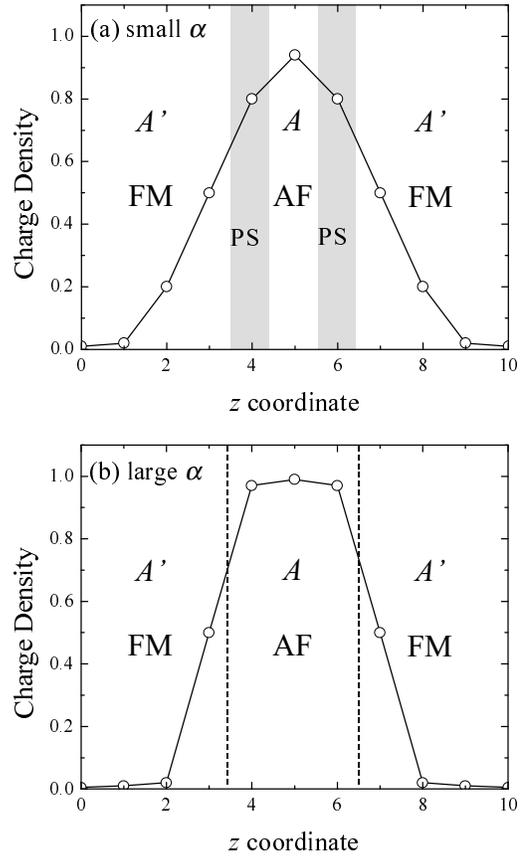}
   \caption{Schematic representation of expected electronic density and low 
   $T$ phase behavior
   of a long period [$m=4, n=6$ heterostructure, with $A$ (charge +1) ions at positions
   $z=3.5, 4.5, 5.5, 6.5$]. The central region is expected to exhibit order
   characteristic of the $n=1$ bulk material (Neel AF in the present case);
   the outer region is ferromagnetic, and the two regions are separated by
   a window of phase separation whose existence depends on the charge binding
   parameter $\alpha$. (a) For weak charge binding (small $\alpha$), there exists phase separated 
   regions between FM and AF states; (b) for strong charge binding (large $\alpha$), 
   no intermediate state separates FM and AF states.}
   \label{fig:speculation}
\end{figure}

We now consider the implications of our results for more general
heterostructures. A given system may be characterized by a charge screening
length $L_{TF}$, which depends on both the charge screening parameter
$\alpha$ and the nature of magnetic order (if any). Systems with 
$L_{TF} > (na, ma)$ exhibit bulk-like behavior with average charge density
$m/(m+n)$; systems with smaller $L_{TF}$ may exhibit spatially differentiated
behavior, with high density and low density regimes characterized by
different kinds of long ranged order. This is seen in the phase diagram shown 
in Fig.\ref{fig:T-alpha phase diagram}, where as $L_{TF}$ is reduced to below
a value of the order of one lattice constant, the central layer exhibits a
different form of long-ranged order than the outer layers. We expect the same
behavior to occur in longer-period structures, with the obvious shifts in phase 
boundaries following from the charges in the length scales to which
$L_{TF}$ should be compared. The resulting two phase structure raises the issue
of the interface between different phases. If $L_{TF}$ is of order one lattice 
constant or less, then we expect an abrupt change of behavior, as is seen
in the $1/\alpha \approx 1$ regime of Fig.\ref{fig:T-alpha phase diagram},
where one layer is AF and the adjacent layers are FM. However, if $L_{TF}$
is larger [but still smaller than ($ma, na$)] then we expect a more gradual 
interface, with one or a sequence of intermediate phases. This behavior is seen
in the ``PS" range ($1.5< 1/\alpha< 2.9$) of Fig.\ref{fig:T-alpha phase diagram}.
Figure \ref{fig:speculation} depicts our expectation of the electronic density and 
the associated phase at each layer for a long period superlattice. AF and FM phases
are separated by a phase separated region whose existence depends on the
screening parameter $\alpha$.
We emphasize that these considerations should apply not only to the specific
double exchange model considered here, but also to other situation in which long
ranged order is controlled by charge density, for example those involving orbital
ordering.

In conclusion, we have used a detailed analysis of a model system to
gain insight into the electronic phase behavior of correlated 
electrons in electronstatically defined heterostructures. We have
shown that the crucial parameter is the strength with electrons 
are bound to the high-density regions, and have distinguished the 
different types of phases which may occur and classified the types of 
interfaces between phases. Our findings also raise the possibility of an 
interesting magneto-electric coupling. Important directions for future
work include applying the ideas introduced here to orbital ordering, and going 
beyond model systems to make predictions for experimental systems.

We thank J. Eckstein, B. Bhattacharya, M. Kawasaki, amd H. Warusawithana
for helpful discussions and DOE ER-046189 and the Columbia MRSEC for support.




\end{document}